\documentclass{aa}  

\usepackage{graphicx,natbib}
\usepackage{txfonts}
\usepackage{color}

\begin{document}

   \title{Detectability of deuterated water in prestellar cores}

   \author{D. Qu\'enard
          \inst{1,2}
          \and
          V. Taquet\inst{3}
          \and
          C. Vastel\inst{1,2}
          \and
          P. Caselli\inst{4}
          \and
          C. Ceccarelli\inst{5,6}
          }

   \institute{Universit\'e de Toulouse, UPS-OMP, IRAP, Toulouse, France\\
         \and
             CNRS, IRAP, 9 Av. colonel Roche, BP 44346, 31028 Toulouse Cedex 4, France\\
        \and
             Leiden Observatory, Leiden University, P.O. Box 9513, 2300-RA Leiden, The Netherlands\\
        \and
             Max Planck Institute for Extraterrestrial Physics, Giessenbachstrasse 1, D-85748 Garching, Germany\\
        \and
             Universit\'e Grenoble Alpes, IPAG, F-38000 Grenoble, France\\
         \and
             CNRS, IPAG, F-38000 Grenoble, France\\
             }

   \date{Written May 18, 2015}

 
  \abstract
   {Water is an important molecule in the chemical and thermal balance of dense molecular gas, but knowing its history throughout the various stages of the star formation is a fundamental problem. Its molecular deuteration provides us with a crucial clue to its formation history. H$_2$O has recently been detected for the first time towards the prestellar core L1544 with the Herschel Space Observatory with a high spectral resolution (HIFI instrument).}
   {Prestellar cores provide the original reservoir of material from which future planetary systems are built, but few observational constraints exist on the formation of water and none on its deuteration before the collapse starts and a protostar forms at the centre. We report on new APEX observations of the ground state 1$_{0,1}$--0$_{0,0}$ HDO transition at 464 GHz towards the prestellar core L1544. The line is undetected, and we present an extensive study of the conditions for its detectability in cold and dense cloud cores.}
   {The water and deuterated water abundances have been estimated using an advanced chemical model simplified for the limited number of reactions or processes that are active in cold regions (< 15 K). In this model, water is removed from the gas phase by freezing onto dust grains and by photodissociation. We use the LIME radiative transfer code to compute the expected intensity and profile of both H$_2$O and HDO lines and compare them with the observations.}
   {The predicted H$_2$O line intensity of the LIME model using an abundance and structure profile, coupled with their dust opacity, is over-estimated by a factor of $\sim$3.5 compared to the observations. We present several \textit{\emph{ad hoc}} profiles that best-fit the observations and compare the profiles with results from an astrochemical modelling, coupling gas phase and grain surface chemistry. The water deuteration weakly depends on the external visual extinction, the external ISRF, and contraction timescale. The [HDO]/[H$_2$O] and [D$_2$O]/[H$_2$O] abundance ratios tend to increase towards the centre of the core up to 25\% and $\sim$8\%, respectively.}
   {Our comparison between observations, radiative transfer, and chemical modelling shows the limits of detectability for singly deuterated water, through the ground-state transitions 1$_{0,1}$-0$_{0,0}$ and 1$_{1,1}$-0$_{0,0}$ at 464.9 and 893.6 GHz, respectively, with both single-dish telescope and interferometric observations. This study also highlights the need of a detailed benchmark amongst different radiative transfer codes for this particular problem of water in prestellar cores.}


   \maketitle
%

\section{Introduction}

Water is an important molecule not only on Earth but also in space. Indeed, since it is formed by the two most abundant elements in the Universe that bind in molecules, water governs the chemical composition and the thermal balance of the interstellar dense molecular gas. This is also the gas from which stars are formed, so that water influences the whole star formation process at various levels in the different phases \citep[e.g.][]{caselli_ceccarelli2012}.
In molecular clouds, iced water is present in large quantities, up to half the oxygen elemental abundance, and is synthesised on the interstellar grains \citep[e.g.][]{boogert2015}. In the denser regions inside the molecular clouds, which are the prestellar cores that will eventually form stars, water is still mostly iced \citep{caselli2012}. The same water, formed in those first two stages, is then found in the gas phase where the dust is warm enough ($\geq 100$ K) in the hot cores, hot corinos, and protostellar molecular shocks \citep[e.g.][]{vandishoeck2014}. It is then again iced and gaseous in the different warm and cold, respectively, zones of the protoplanetary disks, where planets form \citep[e.g.][]{carr2008, podio2013}. Finally, water is the major component of the volatiles in comets \citep[e.g.][]{bockelee-morvan2014}.

In summary, we see water throughout all the stages of the solar-type star forming process up to the leftovers of this process, represented by comets and meteorites. Nevertheless, we do not fully know its whole history: how it evolves from molecular clouds to comet ices and, perhaps, to terrestrial oceans. A crucial aspect in reconstructing the water history is provided by its deuteration by water molecules with one or two deuterium atoms. This is because molecular deuteration is very sensitive to the moment the molecule is formed, the temperature, and also the environment. And since water changes continuously from ice to vapour, it keeps memory of most, if not all, its past formation history \citep[e.g.][]{ceccarelli2014}. So far, the water deuteration has been measured in only a handful of objects, all of them warm: hot cores, hot corinos, and protostellar molecular shocks \citep[e.g.][]{coutens2012, coutens2013, coutens2014, taquet2013b, persson2014}. In cold molecular clouds, only the upper limits of the iced deuterated water exist \citep{dartois2003, parise2003, aikawa2012}. In prestellar cores, no attempt to measure the deuterated water has been published so far.

This article presents the first upper limit on HDO/H$_2$O in a prestellar core, L1544 (Sect. \ref{l1544}), by combining Herschel observations \citep{caselli2012} with new observations obtained at the APEX telescope (Sect. \ref{obs}). We also report a radiative transfer and chemical study of the ortho-H$_2$O and HDO fundamental lines in L1544 (Sects. \ref{water_model} and  \ref{hdo_model}, respectively), and show that the two HDO fundamental lines at 464 and 893 GHz lines are not observable with the present facilities. A final section (Sect. \ref{ccl}) summarises the results.

\section{The prestellar core L1544}\label{l1544}

L1544 is a prototypical starless core in the Taurus molecular cloud complex (d $\sim$ 140 pc) on the verge of gravitational collapse (\citealp{caselli2002a} and references within). It is characterised by high density in its centre (peak density of 2$\times$10$^7$ cm$^{-3}$; \citealp{keto2010}), low temperature ($\sim$ 7 K; \citealp{crapsi2007}), and high CO depletion in its centre, accompanied by a high degree of molecular deuteration (\citealp{caselli2003}; \citealp{crapsi2005}; \citealp{vastel2006}).

Its physical and dynamical structure has recently been reconstructed by \citet{caselli2012} and \citet{keto2014} using the numerous existing observations towards L1544. Among them, the recent detection of water vapour by the Herschel Space Observatory is spectacular because it represents the very first water detection in a prestellar core \citep{caselli2010, caselli2012}. The first of these two Herschel observations was made with the wide-band spectrometer (WBS) with a spectral resolution of 1.1 MHz, and water vapour was detected in absorption against the weak dust continuum radiation ($\sim$ 10 mK) in the cloud. Follow-up observations using the High Resolution Spectrometer (HRS) confirmed the absorption and even detected an inverse P-Cygni emission line profile, too narrow to be seen by the WBS,  which was predicted by theoretical modelling \citep{caselli2010}. 

This detection provided crucial information for reconstructing the physical and chemical structure of L1544. Indeed, this inverse P-Cygni profile, which is characteristic of gravitational contraction, confirmed that L1544 is on the verge of collapsing. Based on the line shape, \citet{caselli2012} predict that water is largely frozen into the grain mantles in the interior ($\leq$~4000 au) of the L1544 core, where the gaseous H$_2$O abundance (with respect to H$_2$) is $< 10^{-9}$. This level of water vapour is believed to be caused by non-thermal desorption processes such as \textit{(a)} the photo-desorption of water molecules from the icy mantles by the far UV photons created locally by the interaction of cosmic rays with H$_2$ molecules and the far UV starlight, and \textit{(b)} the exothermicity of the grain surface chemical reactions that releases the products (e.g. H$_2$O and HDO) in the gas phase \citep{vasyunin2013,wakelam2014}. Farther away from the centre ($\sim10^4$ AU), where the density is low enough ($\leq 10^4$ cm$^{-3}$) for the photo-desorption rate not to be overcome by the freeze-out rate, the gaseous H$_2$O abundance reaches $\sim 1\times 10^{-7}$, in agreement with predictions from comprehensive chemical models \citep{hollenbach2009}.

\section{Observations}\label{obs}

We observed the HDO fundamental transition 1$_{0,1}$--0$_{0,0}$ on April 16, 17, 28, 29, and 30, 2013 and November 2 and 3, 
2013 towards L1544 ($\alpha_{2000} = 05^h04^m17.21^s, \delta_{2000} = 25\degr10\arcmin42.8\arcsec$) using the heterodyne instrument (APEX-3) of the APEX observatory. The frequency was centered at 464.92452 GHz to reach the 1$_{0,1}$--0$_{0,0}$ ground-state HDO transition, and the RPG eXtended bandwidth Fast Fourier Transform Spectrometer (XFFTS) backend was used to obtain the highest frequency resolution needed for a comparison with the H$_2$O profile. To be consistent with the H$_2$O Herschel/HIFI observations, we mapped the HDO transition within the Herschel/HIFI 40$^{\prime\prime}$ beam and reached an rms of 50 mK in about 0.1 km~s$^{-1}$ velocity bin, averaging all positions in order to compare both observations. 
Weather conditions were excellent between 0.2 and 0.7 mm of precipitable water vapour with system temperatures less than 1000 K. 
Line intensities are expressed in units of main-beam brightness temperature with a main beam efficiency of 60\%\footnote{http://www.apex-telescope.org/telescope/efficiency/}.
The ortho-H$_2$O (1$_{10}$-1$_{01}$) line observation was taken from \citet{caselli2012} and was observed with Herschel/HIFI. The dust continuum emission flux at 557 GHz is $10.2\pm0.2$~mK and the rms noise level is 3.8 mK in the spectrum.

\section{H$_2$O modelling}\label{water_model}

\subsection{LIME radiative transfer model}

We ran several radiative transfer modellings of the water emission using LIME, a 3D radiative transfer code \citep{brinch2010} based on ALI (accelerated lambda iteration) calculations. Created from RATRAN-1D \citep{ratran}, LIME does a full radiative transfer treatment in two steps. The first one is to compute the population level of every molecular transition found in the input collision file of the desired molecule. The collisional excitation rates for ortho-H$_2$O are not the same if we consider a collision with ortho-H$_2$ or para-H$_2$. In the case of L1544, we assume that all the hydrogen is in the para state as required by recent chemical models to produce the high deuterium fraction observed in cold, dense clouds \citep{flower2006, pagani2007, troscompt2009, sipila2013, kong2015}. We used the o-H$_2$O - p-H$_2$ collision value from \citet{dubernet2009}, and we assumed a H$_2$O ortho-to-para ratio of 3.

To calculate the population level, LIME needs the 3D structure of the source defined in each point of a model as a function of its cartesian coordinates (X, Y, Z). The source model is usually defined by a few thousand points distributed randomly among a desired radius. Each point is the centre of a 3D cell, and LIME defines the physical properties of the model in each 3D cells (density, temperature, velocity profile, etc.). At least 10~000 points are required to construct the model in order to prevent undersampling. Below 10~000 points, neighbouring 3D cells will be created with different sizes, and non-homogeneous effects will appear in the resulting image after the \textit{ray-tracing}. Thanks to its random distribution, each one of the 10~000 points is located at a unique radius, thus uniquely defined by its physical property. The distribution of points across the model as a function of the radius follows a power law type distribution. This leads to an increasing number of model points per unit volume towards the centre of the cloud, since a finer sampling is needed where the volume densities are higher.

In the second steps, LIME performs a \textit{\emph{ray-tracing}} to output the desired image, depending on the user choice of spatial and spectral resolution and of the number of channels, for instance. As a result, a hyper-spectral cube is created for each chosen line transition. To finish, this cube can be convolved with the beam of the telescope to obtain the spectrum.\\

The large Einstein coefficient (3.45\,$\times$\,10$^{-3}$~s$^{-1}$) of the H$_2$O 1$_{1,0}$--1$_{0,1}$ transition results in optical depths across the core up to a thousand, depending on excitation, leading to a very non-linear relationship between the opacity and the column density. This opacity effect slows down the computation of the population level of the line and can lead to a wrong excitation.
\citet{keto2014} used MOLLIE (\citealp{keto1990}; \citealp{keto_rybicki2010}) to fit their Herschel/HIFI data. For the particular problem of water in L1544, MOLLIE has been modified to treat the radiative transfer with an escape probability method. This method optimises the calculation for the optically thick, but highly sub-thermally excited H$_2$O line towards L1544. Therefore, LIME and MOLLIE are two distinct ways to treat the water modelling problem, and it is necessary to model both H$_2$O and HDO with LIME in order to make a consistent comparison. More information about the differences between MOLLIE and a full radiative transfer code such as LIME will be discussed in a dedicated paper.

\subsection{Grid results}

\begin{figure}[!h]
   \centering
   \includegraphics[width=1\hsize,clip=true,trim=80 20 10 0]{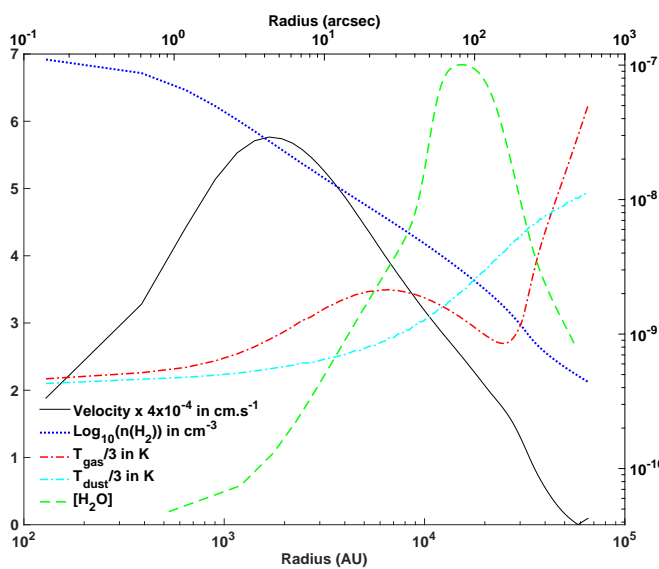}
   \caption{Gas and dust temperature, density, velocity, and water abundance profiles from \citet{keto2014}. The abundance profile is plotted on a logarthmic scale on the right axis of the figure.}
              \label{struct}
    \end{figure}

We used the physical structure as derived by \citet{keto2014} to solve the molecular excitation of the H$_2$O transition. Figure \ref{struct} shows the structure for a slowly contracting cloud in quasi equilibrium. To reproduce the observed line broadening, we considered a Doppler parameter $\beta$ ranging between 100 and 300 m\,s$^{-1}$. We found that a value of 200 m\,s$^{-1}$ gives the best line width fit compared to the observations. This result is consistent with the low gas temperature of L1544.

           \begin{figure}[!h]
   \centering
   \includegraphics[width=1\hsize,clip=true,trim=15 30 0 50]{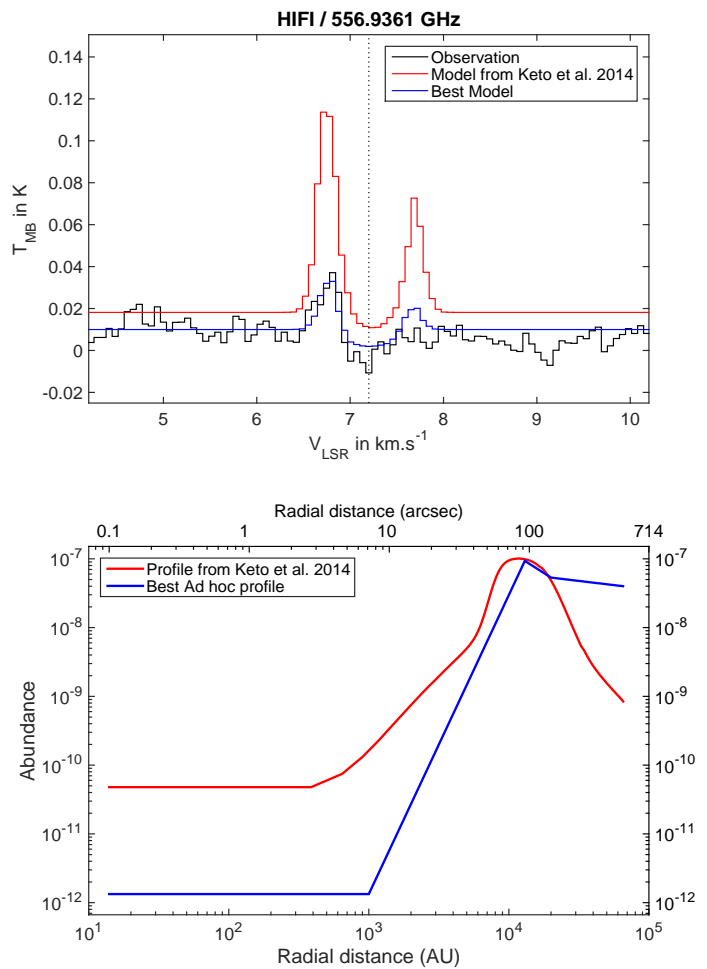}
   \caption{\textit{Top panel:} Comparison between observations and modelling obtained by LIME using the structure and abundance profile from \citet{keto2014} (in red). The best fit model is shown in blue, using the same structure but an \textit{\emph{ad hoc}} abundance profile. \textit{Bottom panel:} Abundance profile from \citet{keto2014} in red and best fit \textit{\emph{ad hoc}} abundance profile in blue as a function of the radial distance from the core in arcsec (top axis) and AU (bottom axis).}
              \label{best_fit}
    \end{figure}

Because of the absorption feature in the H$_2$O observation, it is important to take the continuum value derived from the observations into account and compare its value to the modelling. The LIME code can deal with different input dust opacity files as a function of wavelength such as the tabulated files from \citet{ossenkopf94}. The observed continuum value at 557 GHz (or 538.2 $\mu$m) is 10.2$\pm$0.2 mK \citep[see][]{caselli2012}, and the best model value found with LIME is 10.0 mK with $\kappa_{557}\,=\,8.41$~cm$^{2}$\,g$^{-1}$. \citet{keto2010} based their dust opacity value on the results found by \citet{zucconi2001} where the authors approximate the grain opacities of \citet{ossenkopf94} by piecewise power laws. \citet{keto2010} derived a value of $\kappa_{557}$ = 3.83~cm$^{2}$\,g$^{-1}$ thanks to the following equation describing one of these piecewise power laws \citep[see Appendix B of][]{zucconi2001}:

\noindent
\begin{eqnarray}
    \kappa_\nu = \frac{Q_\nu}{m(H_2)}\times\left(\frac{\lambda_a}{\lambda_{H_2O}}\right)^\alpha\times\left(\frac{m_{gas}}{m_{dust}}\right)~,
        \label{def:kappa}
\end{eqnarray} 

\noindent with $Q_\nu$ = 3.3~$\times$~10$^{-26}$~cm$^2$\,H$_2^{-1}$ at $\lambda_a$ = 1060~$\mu$m, $m(H_2)$ = 3.35~$\times$~10$^{-24}$~g, and $\alpha$ = 2.0. They had to increase the dust opacity value by a factor of four ($\kappa_{557}$ = 15.3~cm$^{2}$\,g$^{-1}$) to be able to reproduce the low temperatures measured by \citet{crapsi2007} towards the centre of L1544. 
    
   \begin{figure*}[!t]
   \centering
   \resizebox{\hsize}{!}
   {\includegraphics[width=1\hsize,clip=true,trim=0 30 10 30]{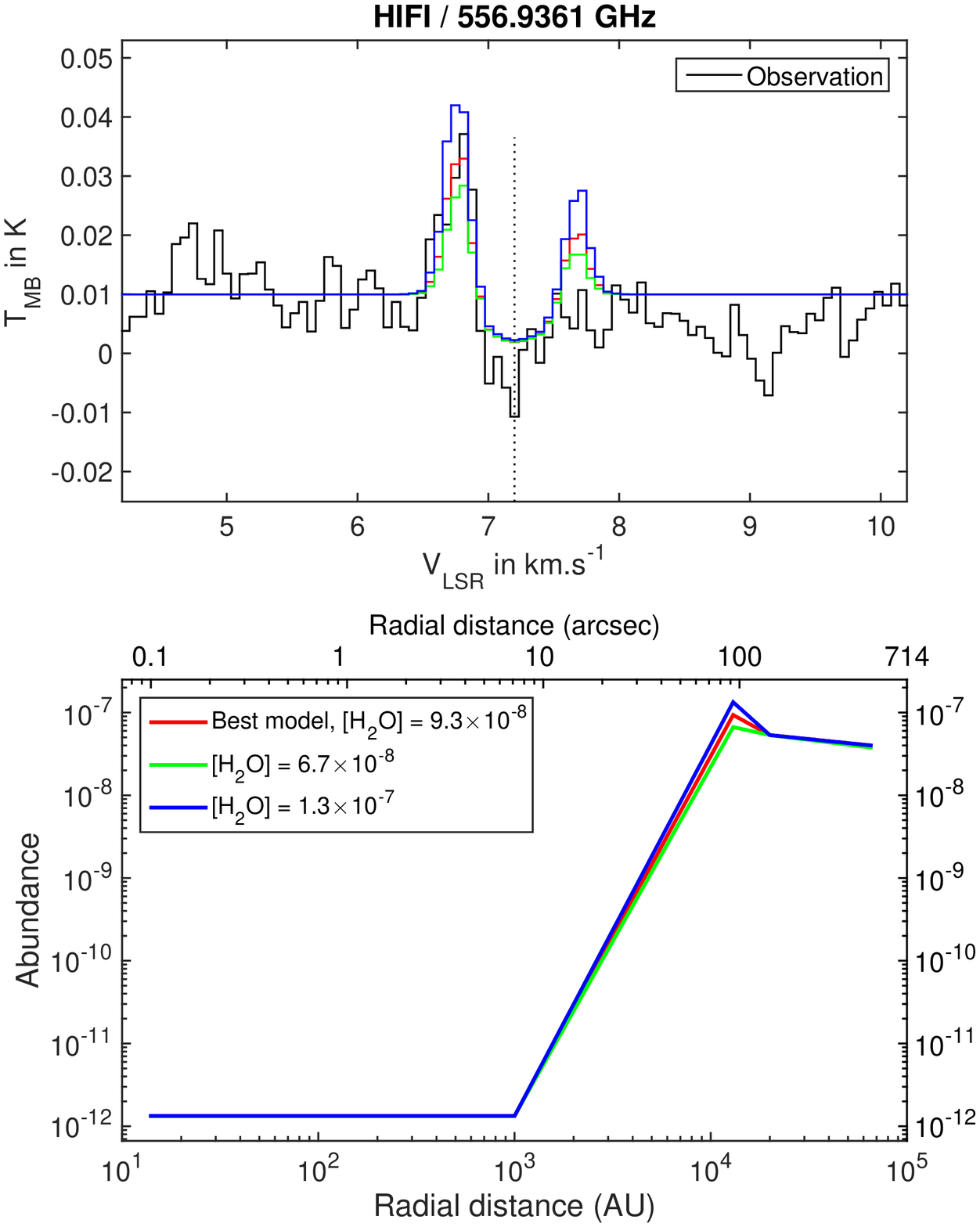}
   \includegraphics[width=1\hsize,clip=true,trim=0 30 10 30]{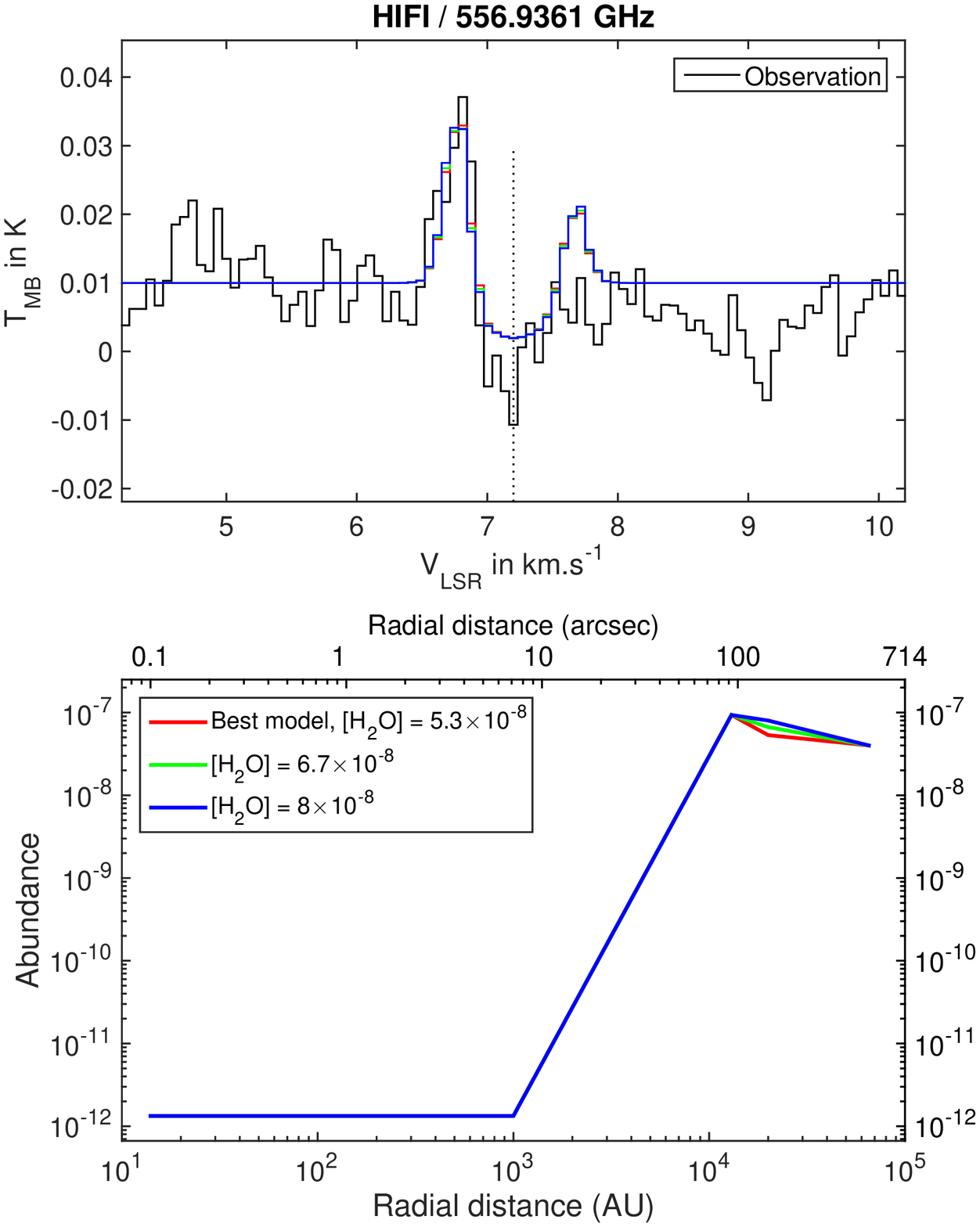}}
   \caption{\textit{Top panels:} Line profile versus observation for three different modellings. The best fit model is shown in red. In the left panel, the abundance value at 13~000 AU varies between $6.7\times10^{-8}$ and $1.3\times10^{-7}$ and for the right panel, the abundance value at 20~000 AU varies between $5.3\times10^{-8}$ and $8\times10^{-8}$. \textit{Bottom panel:} H$_2$O abundance profile used for these respective models as a function of the radial distance from the core in arcsec (top axis) and AU (bottom axis).}
   \label{grid1}
   \end{figure*}
    
The result from the LIME modelling with the abundance and structure profile from \citet{keto2014}, coupled with their dust opacity value, is shown in Fig. \ref{best_fit}. The H$_2$O emission is over-estimated by a factor of $\sim$3.5, and the dust continuum emission is overestimated by a factor of $\sim$2 compared to observations. The latter effect is due to the higher dust opacity ($\kappa_{557}$ = 15.3~cm$^{2}$\,g$^{-1}$) they used compared to the one we found in our best fit ($\kappa_{557}\,=\,8.41$~cm$^{2}$\,g$^{-1}$). As discussed above, the difference in the intensity of the line can be explained by the different radiative transfer treatments in MOLLIE and LIME, including different grids describing the same physical structure, which will be investigated in a forthcoming paper.\\

Although we note that \citet{keto2010} and \citet{keto2014} present a self-consistent model, where a simple chemistry is followed during dynamical and thermal evolution, and where the dynamical evolution is constrained by previous observations of N$_2$H$^+$ and CO isotopologues, we decided here to explore possibly different water abundance profiles to find a good match to the H$_2$O observations using LIME. For this purpose, we modified the water abundance profile found by \citet{keto2014}, while using their physical structure constrained with the many data published for many years towards this source. The inverse P-Cygni profile is a combination of blue-shifted emission and red-shifted absorption, split in velocity by the inward gas motion in front and at the rear of the cloud. In L1544, the emission is shifted with respect to the absorption by less than a line width, leading to an asymmetry in the line profile. Therefore, only a complex modelling can tentatively disentangle between the blue-shifted emission and red-shifted absorption. The emission is produced in the inner region of the cloud, where the density approaches the critical density of the transition. Meanwhile the absorption layer is located in the outer part of the cloud, where the water abundance peaks.

Several \textit{\emph{ad hoc}} profiles have been tested through grid calculations to find the best fit shown in Fig. \ref{best_fit}, varying the following different parameters located in four distinct regions of the core:
\begin{enumerate}
        \item The first region is the inner abundance and we found a best fit value of [H$_2$O] = $1.3\times10^{-12}$ at the centre of the core until a radius of 1000 AU. \citet{keto2014} found a value of this inner abundance close to [H$_2$O] = $5\times10^{-11}$. If we increase the inner abundance to this value, we also change the slope of the abundance profile, probing a region where the velocity field is higher. This effect leads to a wider and slightly blue-shifted emission compared to our best fit. Nonetheless, due to the large HIFI beam at 557 GHz, the constraints on the inner abundance are relatively poor. Thus, the contribution of cosmic-ray-induced UV photons responsible for this inner abundance is also poorly constrained, and a proper benchmark between LIME and other radiative transfer codes is needed to develop a deeper analysis of this inner abundance. Cosmic-ray-induced UV photons also appear important for reproducing the observations of CO isotopologues toward L1544 \citep{keto2010}.
        \item The second region defines the peak abundance value of the external layer. We varied the abundance value and the distance from the centres of the core of this region. We found a best fit value of [H$_2$O] = $9.3\times10^{-8}$ at a radius of 13~000 AU.
        \item The third region marks the end of the external layer and, along with the first region, we varied the abundance value and the distance from the centre of the core. We found a best fit value of [H$_2$O] = $5.3\times10^{-8}$ at a radius of 20~000 AU.
        \item We finally varied the abundance value at the end of the profile, located at the same final radius of the physical structure of the core given by \citet{keto2014}. We found a best fit value of [H$_2$O] = $4\times10^{-8}$. This result may imply that L1544 is well embedded in a relatively large filament in Taurus, so that the extinction $A_{\textrm{V}}$ is high enough to at least partially shield interstellar UV photons (see Section \ref{chem}).
\end{enumerate}

We decided to first determine the first and second regions at the same time thanks to the grid. We then fixed the third one, which appeared to be independent of the other two.
We derived the best fit from these previous models with the help of the standard $\chi^2$ minimization value for a spectrum of N points, given by the equation \citep{coutens2012}:
\begin{eqnarray}
    \chi^{2}~=~\sum_{j=1}^{N} \dfrac{(I_{obs,j} - I_{model,j})^{2}}{rms^{2} + (cal\times I_{obs,j})^{2}},
        \label{def:chi2}
\end{eqnarray} 
where $I_{obs}$ and $I_{model}$ are the observed and the modelling intensity, $cal_{i}$ is the calibration factor of the spectrum, and $N$ is the total number of points of the spectrum. In the left-hand panels of Fig. \ref{grid1} we show a small variation in the peak abundance value at the best fit distance of 13~000 AU from the centre of the core. It is interesting to note that a little variation in the abundance value at this distance can cause a notable difference in the line profile. Meanwhile, if we change the abundance value at the best fit distance of 20~000 AU of the external layer (see right panels of Fig. \ref{grid1}), the emission is the same for the three profiles.

In fact, emission from a molecule like water with high critical densities is only possible in regions with sufficiently high H$_2$ densities. In L1544, this region corresponds to the back part of the core approaching the centre (thus moving towards us) and revealing infall through an inverse P-Cygni profile (emission in the blue-shifted part of the line). We are seeing differences in the height of the blue peak by changing the abundance of water around 13~000 AU owing to more or fewer water molecules being present in the outer envelope to be able to absorb the emission coming from the central regions. This results does not contradict the detection of methanol and complex organic molecules (COMs) at radii of about 10~000 AU from the centre \citep{vastel2014, bizzocchi2014}, because water, methanol, and COMs have to be present in the gas phase at such radii. However, water at the volume densities present in the outer envelope can only absorb, while methanol and other complex molecules can emit more easily (as their critical densities are not as high). If the water abundance value at the edge of the core is lower than [H$_2$O] $\simeq 1\times10^{-8}$, the absorption feature will not be deep enough.

\subsection{Chemical modelling of H$_2$O and HDO}\label{chem}

We modelled the abundance profile of standard and deuterated water in
L1544 using the GRAINOBLE astrochemical model, described in \citet{taquet2012,taquet2014}. Briefly, GRAINOBLE couples the gas phase and grain
surface chemistry with the rate equation approach introduced by \citet{hasegawa1992} during the static contraction of a starless
core.
The gas-phase chemical network was taken from the KIDA database and
has been extended to include the spin states of H$_2$, H$_2^+$,
H$_3^+$, and the deuterated isotopologues of hydrogenated species with
four or fewer atoms among with species involved in the gas phase
chemical network of water, ammonia, formaldehyde, and methanol. A more
detailed description of the chemical network is presented in \citet{taquet2014}.
We also considered the following gas-grain processes:

\begin{enumerate}
\item Accretion of gas phase species on the surface of spherical grains
with a fixed diameter $a_{\textrm{d}}$ assumed to be equal to 0.1
$\mu$m.
\item Diffusion of adsorbed species via thermal hopping, exponentially
depending on the diffusion-to-binding energy ratio
$E_{\textrm{d}}/E_{\textrm{b}}$. We set $E_{\textrm{d}}/E_{\textrm{b}} = 0.65$, following our previous studies.
\item Reaction between two particles via the Langmuir-Hinshelwood
mechanism.
\item Desorption of adsorbed species into the gas phase by thermal
evaporation, interstellar plus cosmic-ray induced heating of grains, chemical
evaporation, and UV
photolysis. The thermal evaporation exponentially depends on the
binding energy of each species $E_{\textrm{b}}$ relative to the
substrate \citep[see][for a list of binding energies used in the model]{taquet2014}. The cosmic-ray-induced heating of grains follows the
approach by \citet{hasegawa1993a} and is adapted to the binding
energies considered in this work. The approach adopted for the UV photolysis
follows the method described in \citet{taquet2013a}.
\end{enumerate}

We used the multi-layer approach developed by \citet{hasegawa1993b} to follow the multi-layer formation of interstellar ices and
considered three sets of differential equations: one for gas-phase
species, one for surface species, and one for bulk species.
The equations governing chemical abundances on the surface and in the
bulk are linked by an additional term that is proportional to the rate
of growth or loss of the grain mantle. As a consequence,
surface species are continuously trapped in the bulk because of the
accretion of new particles. 

We followed the formation and the deuteration of the main ice species
following the surface chemical network developed by \citet{taquet2013a}, which is based on laboratory experiments showing the
efficient formation of interstellar ice analogues at low
temperatures. Transmission probabilities for key reactions involved in
the water chemical network have been estimated through quantum
chemistry. 

The gas-grain chemistry is followed during the static contraction of a
dense core starting from a homogeneous translucent sphere of a
initial density $n_{\textrm{H,ini}} = 3 \times 10^3$ cm$^{-3}$ and a
maximal radius of $3\times10^4$ AU. 
During the static contraction, the core keeps a Plummer-like density
profile:
\begin{equation}
n_{\textrm{H}} = \frac{n_{\textrm{H,0}}}{(1+(r/R_{\textrm{f}})^2)^{\eta/2}},
\end{equation}
where $n_{\textrm{H,0}}$ is the central density, and $R_{\textrm{f}}$ 
the characteristic radius inside which the density is
uniform. The contraction ends when the density profile reaches the
observed profile of L1544, with the following parameters: 
$n_{\textrm{H,0}} = 1.8 \times 10^7$ cm$^{-3}$, $R_{\textrm{f}} = 450$ AU, and $\eta =
2.1$.
Since $R_{\textrm{f}}$ is given by the product of the sound speed and the
free-fall time of the central density, $R_{\textrm{f}}$ decreases with
$1/\sqrt{n_{\textrm{H,0}}}$.
Intermediate central densities, and the associated timescale needed
to reach them, have been chosen to have a total contraction
timescale of about one million years, following observational
estimates of molecular cloud cores \citep[e.g.][]{brunken2014}.

\begin{figure}[htp]
\centering
\includegraphics[width=\columnwidth]{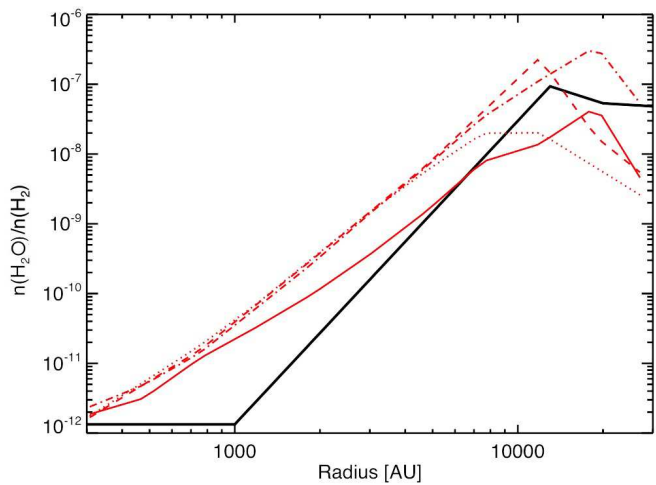}
\caption{Water abundance profile in L1544 modelled with GRAINOBLE (red curves) compared with the observed profile based on LIME radiative transfer simulations (black). Dashed-dotted line: $G_0 = 1$, $A_{\textrm{V,ext}} = 2$, $t_{\textrm{c}} = 7.5 \times 10^5$ yr; dashed line: $G_0 = 1$, $A_{\textrm{V,ext}} = 1$, $t_{\textrm{c}} = 7.5 \times 10^5$ yr; dotted line: $G_0 = 0.1$, $A_{\textrm{V,ext}} = 1$, $t_{\textrm{c}} = 7.5 \times 10^5$ yr; solid line: $G_0 = 1$, $A_{\textrm{V,ext}} = 2$, $t_{\textrm{c}} = 1.5 \times 10^6$ yr.}
\label{Xgrainoble}
\end{figure}

\begin{figure}[htp]
\centering
\includegraphics[width=\columnwidth]{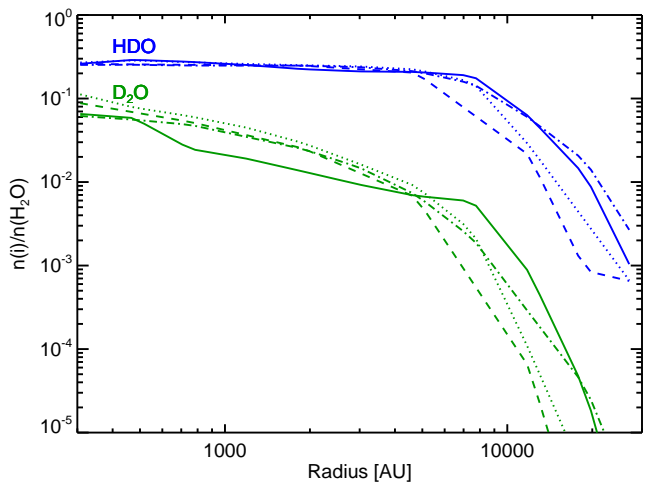}
\caption{[HDO]/[H$_2$O] (blue) and [D$_2$O]/[H$_2$O] (green) deuteration profile in L1544 modelled with GRAINOBLE. Dashed-dotted line: $G_0 = 1$, $A_{\textrm{V,ext}} = 2$, $t_{\textrm{c}} = 7.5 \times 10^5$ yr; dashed line: $G_0 = 1$, $A_{\textrm{V,ext}} = 1$, $t_{\textrm{c}} = 7.5 \times 10^5$ yr; dotted line: $G_0 = 0.1$, $A_{\textrm{V,ext}} = 1$, $t_{\textrm{c}} = 7.5 \times 10^5$ yr; solid line: $G_0 = 1$, $A_{\textrm{V,ext}} = 2$, $t_{\textrm{c}} = 1.5 \times 10^6$ yr.}
\label{Dgrainoble}
\end{figure}

We used the radiative transfer code DUSTY \citep{dusty} to
compute the temporal evolution of the dust temperature profile of the
contracting core by assuming that gas and dust temperatures are
coupled. As shown in Fig. \ref{struct}, the dust and gas temperatures are
decoupled and can differ by up to 4 K. We checked at posteriori that
variations in gas phase temperatures of 4 K only induce small variations in
abundances of gaseous species by 30 \% at most.
The thermal structure of the core is derived from a slab geometry in
which the core is irradiated by the interstellar radiation field (ISRF) with a
spectrum taken from \citet{evans2001} and assuming a fixed
temperature at the edge of core of 13 K following the observed
temperature profile of L1544. 

As discussed in \citet{taquet2013a, taquet2014}, the abundance and the
deuteration of the main ice components like water are known to depend
on various physical and chemical parameters that are either poorly
constrained or that show distributions of values. 
To reproduce the water abundance profile deduced by the LIME radiative transfer study, we varied
the values of three poorly constrained physical parameters that are
thought to have a strong impact on the abundance of gaseous water:

\begin{enumerate}
\item The external visual extinction $A_{\textrm{V,ext}}$ that influences the radius where the water 
abundance reaches its maximal value. As shown in Fig.
\ref{Xgrainoble}, decreasing $A_{\textrm{V,ext}}$ from two to one magnitude
enhances the photodissociation of water at the edge of the core, shifting its
maximal abundance toward the core centre from $1.8 \times 10^4$ to
$1.2 \times 10^4$ AU.  
\item The external ISRF (interstellar radiation field) $G_0$. The photo-desorption rate of water increases
with the flux of external UV photons, which is proportional to $G_0$. The
maximal abundance of water starts to increase with $G_0$ at low $G_0$
and then decreases when the photo-dissociation of gaseous 
water overcomes its photo-desorption. Decreasing $G_0$ from 1 to 0.1
decreases the maximun water abundance from $2 \times 10^{-7}$ to $2
\times 10^{-8}$.
\item The contraction timescale $t_{\textrm{c}}$. The core contraction
timescale impacts the total number of particles that freeze-out on
grains at the centre of the dense core but also the number of
particles that are photo-evaporated at lower densities and visual
extinctions in the outer shells. The increase in the core contraction
timescale between $7.5 \times 10^5$ and $1.5 \times 10^6$ yr decreases
the water abundance in the dense part of the core, owing to higher depletion,
but also slightly increases the water abundance towards the edge
because of the higher total number of photo-evaporated water molecules.
\end{enumerate}

The water deuteration profiles obtained for the three sets of parameters used
to model the water abundance profile in Fig. \ref{Xgrainoble} are
shown in Fig. \ref{Dgrainoble}.
The water deuteration weakly depends on the physical parameters: the
[HDO]/[H$_2$O] and [D$_2$O]/[H$_2$O] abundance ratios tend to increase
towards the centre of the core up to $\sim 25$ \% and $\sim 8$ \%,
respectively. As comprehensively studied in previous analyses \citep{roberts2004, flower2006, taquet2014, sipila2015}, 
the increase in the deuteration both in the gas phase and on
ices toward the core centre is due to the decrease in both the CO gas phase
abundance and the H$_2$ ortho/para ratio, the two main parameters
limiting the deuterium chemistry, with the increase in the total
density and the decrease in the temperature.
As also discussed in \citet{taquet2014}, the gas phase deuteration
of water obtained at the centre of dense cores is higher by more than
one order of magnitude than the overall deuteration predicted in
interstellar ices. The gas phase D/H abundance ratio of water reflects
the gas phase chemistry and surface chemistry in the outermost ice layers, in interaction
with the gas phase, that are occurring in dense and cold conditions.
The low deuteration of water ice is due to its early formation
in the molecular cloud phase when the CO abundance and the
H$_2$ ortho/para ratio were high.

Finally, the parameters of the chemical model that best fit the LIME-deduced water abundance profile are
$G_0 = 1$, $A_{\textrm{V,ext}} = 2$, and $t_{\textrm{c}} = 1.5 \times 10^6$ yr for a fixed diameter of grain $a_{\textrm{d}}$ assumed to be equal to 0.1
$\mu$m (see Fig. \ref{Xgrainoble}).

\section{HDO modelling with LIME}\label{hdo_model}

       \begin{figure}[ht]
   \centering
   \includegraphics[width=1\hsize,clip=true,trim=0 0 0 0]{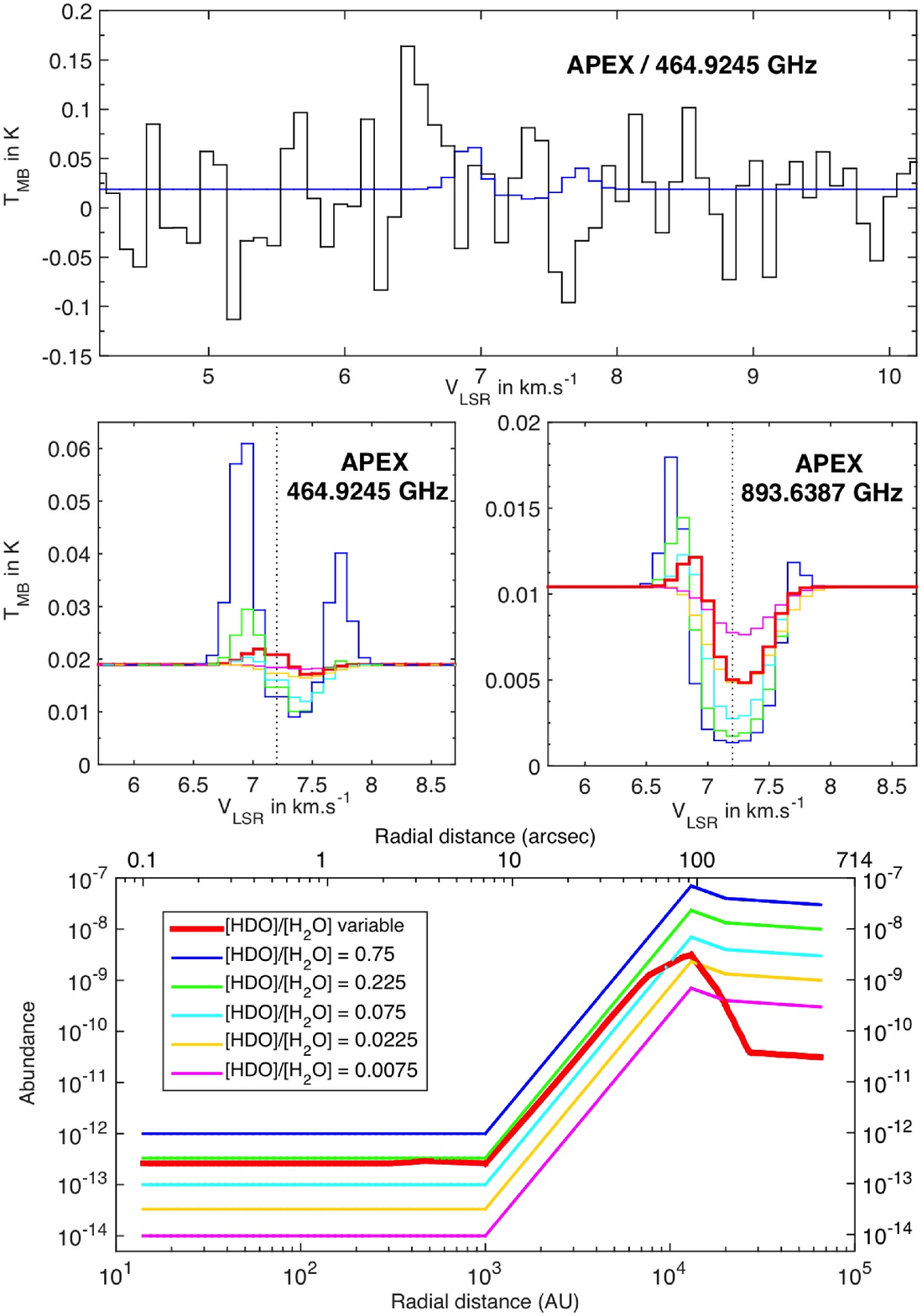}
   \caption{\textit{Top panel:} HDO 1$_{0,1}$--0$_{0,0}$ line profile of the D/H = 1 model (in bue) versus APEX observation (in black). \textit{Middle panels:} HDO 1$_{0,1}$--0$_{0,0}$ (left panel) and 1$_{1,1}$--0$_{0,0}$ (right panel) line profiles for the six different D/H ratios. \textit{Bottom panel:} HDO abundance profiles as a function of the D/H ratio (see text, Section \ref{hdo_model}) used for these models as a function of the radial distance from the core in arcsec (top axis) and AU (bottom axis).}
              \label{HDO_multi_ratio}
    \end{figure}

\citet{coutens2012} have shown that in the outer part of the envelope of the low-mass protostar IRAS16293-2422, the [HDO]/[H$_2$O] ratio is about 0.2-2.2\% and $\sim$4.8\% in the external layer. This layer can be associated to the parental cloud, a remnant of the initial prestellar core of the protostar. Thus, we can say that an expected realistic [HDO]/[H$_2$O] ratio in L1544 may be close to these values. Nonetheless, observed D/H ratios for other gaseous species, as well as astrochemical models of starless cores, suggest much higher values, ranging from 7.5\% to 22.5\%. We chose to consider six different D/H ratios:
\begin{itemize}
        \item 0.75\% and 2.25\%, in agreement with the [HDO]/[H$_2$O] ratios found by \citet{coutens2012} in the external layer of IRAS16293-2422, a low-mass protostar.
        \item 7.5\% and 22.5\%, in agreement with other observed D/H ratios in molecules, such as HCO$^+$, N$_2$H$^+$, and H$_2$CO in prestellar cores \citep[e.g.][]{bacmann2003, pagani2007, kong2015}.
        \item variable as a function of the radius. We took the D/H ratio found by the dotted astro-chemical modelling plotted in Fig. \ref{Dgrainoble}. This chemical modelling corresponds to $G_0 = 0.1$, $A_{\textrm{V,ext}} = 1$, and $t_{\textrm{c}} = 7.5 \times 10^5$ yr.
        \item 75\% to consider an extreme case where the [HDO]/[H$_2$O] ratio is very high.
\end{itemize}
Based on our best-fit \textit{\emph{ad hoc}} H$_2$O abundance profile, we derived the abundance profile for the six previous different D/H ratios we consider. We modelled the HDO 1$_{0,1}$--0$_{0,0}$ transition at 464.914 GHz and the 1$_{1,1}$--0$_{0,0}$ transition at 893.639 GHz. The APEX observation of the 464.914 GHz transition does not show any detection of HDO towards the source (as seen in the upper panel of Fig. \ref{HDO_multi_ratio}), and the 893.639 GHz transition has not been observed. The predicted continuum value at 464.914 GHz and 893.639 GHz are 18.9 mK and 10.4 mK, respectively. The predicted continuum value is lower at a higher frequency, which is consistent with a cold prestellar core such as L1544. The two middle panels of Fig. \ref{HDO_multi_ratio} show the spectra of the 1$_{0,1}$--0$_{0,0}$ and 1$_{1,1}$--0$_{0,0}$ transition derived from the output brightness map of the six different models, and the lower panel displays the related HDO abundance profile. The upper panel compares the APEX observations of the 464.914 GHz transition and the D/H = 1 model. This comparison clearly shows the limits of detectability for deuterated water with single-dish telescopes such a APEX, even if we consider a very high and unrealistic ratio. Table \ref{abs_peak} shows the intensity of the emission and absorption feature for every model with respect to the continuum level of the line.

\begin{table}
\caption{\label{abs_peak}Intensity of the emission and absorption feature of the two transitions with respect to their continuum level.}
\centering
\begin{tabular}{cc|c||c|c}
\hline\hline
D/H ratio&\multicolumn{2}{c||}{HDO (1$_{0,1}$--0$_{0,0}$)}&\multicolumn{2}{c}{HDO (1$_{1,1}$--0$_{0,0}$)}\\
(in \%)&\multicolumn{2}{c||}{464.914 GHz}&\multicolumn{2}{c}{893.639 GHz}\\
\hline
&absorption&emission&absorption&emission\\
&(in mK)$\tablefootmark{b}$&(in mK)&(in mK)$\tablefootmark{b}$&(in mK)\\
\hline
variable$\tablefootmark{a}      $&1.9&2.9&5.4&1.5\\
75\%&9.9&42.1&9.0&4.8\\
22.5\%&8.8&10.6&8.7&1.5\\
7.5\%&6.9&1.4&7.7&1.3\\
2.25\%&2.5&0.04&5.5&0.04\\
0.75\%&0.8&0.007&2.6&0.002\\
\hline\hline
\end{tabular}
\tablefoot{
\tablefoottext{a}{D/H ratio as a function of the radius, derived from the dotted astro-chemical modelling shown in Fig. \ref{Dgrainoble}.}
\tablefoottext{b}{Absorption with respect to the continuum value of 18.9 mK at 464.914 GHz and 10.4 mK at 893.639 GHz.}
}
\end{table}

By looking at the brightness map of any of the previous models, one can note that the absorption feature of the HDO lines only shows up on a 50\arcsec~scale around the centre of the model. If we consider the most compact Cycle 3 configuration of ALMA (C36-1), it is impossible to entirely map L1544 in a convenient number of pointings owing to the size of the core ($\sim20~000$ AU, which means $\sim143\arcsec$ at 140 pc), and even if we consider only the absorption feature ($\sim50\arcsec$), it would still require too many pointings since the antenna beamsize of the C36-1 configuration is $\sim12.5\arcsec$ and $\sim6.5\arcsec$ for the two HDO transitions. The ALMA interferometer is not optimum here to detect a very extended emission, and complementary ACA observations will not be sufficient to cover the missing (u,v)-plane observations. Only a single-dish telescope can give a convenient beam size to cover at least the absorption feature of L1544, but right now, such low sensitivities ($\sim2$ mK at 464.914 GHz and $\sim5$ mK at 893.639 GHz) cannot be reached with ground-based observatories, and no space observatory is foreseen in the future.

\section{Conclusions}\label{ccl}

 Based on the recent detection of the 1$_{1,0}$--1$_{0,1}$ water transition using the Herschel/HIFI instrument, we used the APEX observatory to constrain the water fractionation in the L1544 prestellar core.
        We used LIME to model the 1$_{1,0}$--1$_{0,1}$ H$_2$O and 1$_{0,1}$--0$_{0,0}$ HDO line profiles towards L1544 with a full radiative transfer treatment in 3D.
         \citet{keto2014} derived the density, temperature, and velocity profile of the source. However, using LIME instead of MOLLIE, we found that their abundance profile predicts an emission that is 3.5 times stronger than the observation while using their deduced dust opacity, LIME predicts a continuum emission at 557 GHz about two times stronger than the observed value.

This result points to the need for a detailed comparison between these codes for the specific case of water in prestellar cores.
         We have found an \textit{\emph{ad hoc}} abundance profile that fits the Herschel/HIFI H$_2$O observation better using LIME. We also found a new estimation of the dust opacity to reproduce the observed continuum of $\sim$ 10.2 mK.
         We used a detailed chemical modelling using both gas-phase and grain-surface chemistry to reproduce the water profile and predict its subsequent deuterated water profile. The resulting profile ($G_0 = 1$, $A_{\textrm{V,ext}} = 2$, $t_{\textrm{c}} = 1.5 \times 10^6$ yr, $a_{\textrm{d}} = 0.1$ $\mu$m) has then been compared with our APEX observations. Our study shows the limit of detectability for deuterated water in prestellar cores using ground-based facilities with both single-dish telescopes and interferometric antennas.

\begin{acknowledgements}
The authors are grateful to Carlos De Breuck for his help during the APEX observations. C.C. acknowledges the financial support by the French Space Agency CNES. P.C. acknowledges the financial support of the European Research Council (ERC; project PALs 320620). 
\end{acknowledgements}


\end{document}